\documentclass[12pt]{article}
\usepackage{osajnl2}

\begin{document}

\newcommand{\sqvb}{\ensuremath{ \langle \!\langle 0 |} }
\newcommand{\sqvk}{\ensuremath{ | 0 \rangle \!\rangle } }
\newcommand{\sqvn}{\ensuremath{ \langle \! \langle 0 |  0 \rangle \! \rangle} }
\newcommand{\wh}{\ensuremath{\widehat}}
\newcommand{\be}{\begin{equation}}
\newcommand{\ee}{\end{equation}}
\newcommand{\bea}{\begin{eqnarray}}
\newcommand{\eea}{\end{eqnarray}}
\newcommand{\ra}{\ensuremath{\rangle}} 
\newcommand{\la}{\ensuremath{\langle}}
\newcommand{\rra}{\ensuremath{ \rangle \! \rangle }}
\newcommand{\lla}{\ensuremath{ \langle \! \langle }}
\newcommand{\str}{\rule[-.125cm]{0cm}{.5cm}}
\newcommand{\pr}{\ensuremath{^\prime}}
\newcommand{\ppr}{\ensuremath{^{\prime \prime}}}
\newcommand{\da}{\ensuremath{^\dag}}
\newcommand{\as}{^\ast}
\newcommand{\eps}{\ensuremath{\epsilon}}
\newcommand{\ve}{\ensuremath{\vec}}
\newcommand{\ka}{\kappa}
\newcommand{\non}{\ensuremath{\nonumber}}
\newcommand{\lf}{\ensuremath{\left}}
\newcommand{\rt}{\ensuremath{\right}}
\newcommand{\al}{\ensuremath{\alpha}}
\newcommand{\dfn}{\ensuremath{\equiv}}
\newcommand{\ga}{\ensuremath{\gamma}}
\newcommand{\ti}{\ensuremath{\tilde}}
\newcommand{\wti}{\ensuremath{\widetilde}}
\newcommand{\hs}{\ensuremath{\hspace*{.5cm}}}
\newcommand{\bet}{\ensuremath{\beta}}
\newcommand{\om}{\ensuremath{\omega}}

\newcommand{\cO}{\ensuremath{{\cal O}}}
\newcommand{\cS}{\ensuremath{{\cal S}}}
\newcommand{\cF}{\ensuremath{{\cal F}}}
\newcommand{\cX}{\ensuremath{{\cal X}}}
\newcommand{\cZ}{\ensuremath{{\cal Z}}}
\newcommand{\cG}{\ensuremath{{\cal G}}}
\newcommand{\cR}{\ensuremath{{\cal R}}}
\newcommand{\cV}{\ensuremath{{\cal V}}}
\newcommand{\cC}{\ensuremath{{\cal C}}}
\newcommand{\cP}{\ensuremath{{\cal P}}}
\newcommand{\pup}{\ensuremath{^{(p)}}}
\newcommand{\prpr}{\ensuremath{\prime \prime }}

\OSAJNLtitle{\bf 
Signal-to-Noise Ratio in Squeezed-Light Laser Radar
}
\OSAJNLauthor{
Mark A. Rubin and Sumanth Kaushik}
\OSAJNLaddress{Lincoln Laboratory\\ 
Massachusetts Institute of Technology\\  
244 Wood Street\\                         
Lexington, Massachusetts 02420-9185}
\OSAJNLemail{\{rubin,skaushik\}@LL.mit.edu}

\begin{center}
\begin{abstract*}
The formalism for computing the signal-to-noise ratio (SNR) for laser radar
is reviewed and applied to the tasks of target detection, direction-finding, and
phase change estimation with squeezed light.
The SNR for heterodyne detection of coherent light using a squeezed
local oscillator is lower than that obtained using a coherent local oscillator.
This is true  for target detection,  for phase estimation, and   for direction-finding 
with a split detector. Squeezing the local oscillator also lowers SNR in balanced homodyne and heterodyne detection of coherent light. 
Loss places an upper bound on the improvement that squeezing can bring to  direct-detection SNR.

\end{abstract*}
\end{center}

\maketitle

\section{Introduction}

Squeezed light was proposed in \cite{Caves1981}  
as a means of reducing noise
in interferometers used as gravitational-wave detectors.  In that article 
the impact of loss on performance improvement due to squeezing was noted. In laser radar applications, loss from spreading and absorption in the target-return beam can be severe \cite{Kingston1978}.
It is therefore natural to investigate the potential of using squeezed light in the local oscillator of a heterodyne laser radar system \cite{Lietal1997,Lietal1999,RubinKaushik2007}, since the loss suffered by the local oscillator beam  will be minimal.  Here we review relevant signal detection theory and quantum optics formalism, and compute the signal-to-noise ratio (SNR) for several application scenarios. In all the cases we examine we find that squeezed light offers no advantages for laser radar.

This paper is organized as follows. Sec. \ref{Definitions} and \ref{SecQhsm} review relevant aspects of signal detection theory. Sec. \ref{SecOperators} obtains explicit forms for quantum operators that will be used in the subsequent analysis, and Sec. \ref{Signal}
shows how they fit into the heterodyne detection approach. The SNR for heterodyne target detection using a squeezed local oscillator is
derived in Sec. \ref{SecHettargdet} and is shown to be lower than that obtainable with heterodyne target
detection with the usual coherent local oscillator.
The expression for SNR is given in term of conventional parameters in Sec. \ref{SecSNRconvparams} 
and is extended to include nonunity quantum efficiency in Sec. \ref{SecQuanteff}.
Sec. \ref{SecDirfind}  presents and analyzes models of heterodyne and direct-detection direction-finding using a split detector,
and shows that here too squeezing the local oscillator of the heterodyne system yields no advantage.
Sec. \ref{SecHetphase} looks at heterodyne detection of phase. Sec. \ref{SecBaldet} extends the result of
Sec. \ref{SecHettargdet} to the case of balanced detection. The Appendix examines direct target detection with
squeezed light and derives limits on  SNR improvement.

\section{Signal-to-noise ratio} \label{Definitions}

From \cite{Helstrom1976}, Ch. IV, ``Quantum Hypothesis Testing:''

``Suppose that when the signal is present, it is repeated during some number $M$\/
of observation intervals of duration $T$\/. Let $\vec{x}_k$\/ be a set of
data samples taken in the $k^{th}$\/ interval, $k=1,2,\ldots,M$\/.  We assume these are
statistically independent from one another.  Let $g(\vec{x}_k)$\/ be a statistic formed from the data
$\vec{x}$\/, and let the choice between hypotheses $H_0$\/ and $H_1$\/ be based on
the sum
$$
\hspace*{2.5in}G=\sum_{k=1}^M g(\vec{x}_k)\hspace*{2.5in}\mbox{(3.2)}
$$
of the statistics for each interval\ldots Hypothesis $H_1$\/ is selected if $G$\/
exceeds a certain decision level $G_0$\/.

``When $M$\/ is very large, the p.d.f. [probability density function] of the statistic $G$\/
is very nearly Gaussian, by virtue of the central limit theorem,
and the false alarm and detection probabilities are approximately 
$$
\hspace*{1.6in}Q_0\approx \mbox{erfc}(\xi), \hspace{5mm}Q_d\approx \mbox{erfc}(\xi-D_gM^{1/2})\hspace*{1.6in} \mbox{(3.3)}
$$
where
$$
\hspace*{2in}D_g^2=\frac{[E(g|H_1)-E(g|H_0)]^2}{\mbox{Var}_0g}\hspace*{2in}\mbox{(3.4)}
$$
with 
$$
\hspace*{1.9in}\mbox{Var}_0g=E(g^2|H_0)-[E(g|H_0)]^2\hspace*{1.9in}\mbox{(3.5)}
$$
the variance of the statistic $g$\/  under hypothesis $H_0$\/.
[$E(g|H_i)$\/, $i=1,2$\/, is the ``expected value'' or ``mean value'' \cite{Meyer1970} of $g$\/ when 
hypothesis $H_i$\/ holds.] 
We call $D_g^2$\/ the equivalent signal-to-noise ratio (e.s.n.r.).''

For direct detection $\mbox{Var}_0g=0$\/, so $\mbox{Var}_1g$\/, the variance of the
signal when it is present (i.e., when $H_1$\/ holds) will be used. See Appendix. 

\section{Quantum hypotheses and statistical measures}\label{SecQhsm}

When hypothesis $H_i$\/ holds, the quantum state is $|\psi_i\ra$, $i=0,1$\/.
The mean value of S when $H_i$\/ holds is 
\be
E(S|H_i)=\la \psi_i|\wh{S}|\psi_i\ra,\label{qmeandef}
\ee
where $\wh{S}$\/ is the operator corresponding to the quantity $S$\/. The variance of
$S$\/ when $H_i$\/ holds is
\be
\mbox{Var}_iS=\la\psi_i |\wh{S}^2|\psi_i\ra- \la \psi_i |\wh{S}|\psi_i\ra^2.\label{qvardef}
\ee
Using (\ref{qmeandef}) and (\ref{qvardef}) in (3.4) and (3.5) of Sec. \ref{Definitions}, and taking $g=S$,
\be
D_S^2=\frac{\left(\la \psi_1|\wh{S}|\psi_1\ra-\la \psi_0|\wh{S}|\psi_0\ra\right)^2}{\la \psi_0|\wh{S}^2|\psi_0\ra-\la \psi_0|\wh{S}|\psi_0\ra^2}.\label{DS}
\ee

\section{Operators}\label{SecOperators}

The positive-frequency part of the electric-field operator is
\be
\wh{E}^{(+)}_\mu(t)=\sum_{\vec{k},\zeta}i\left(\frac{\hbar \omega_{\vec{k}}}{2 \varepsilon_0 V}\right)^{1/2}
\wh{a}_{\vec{k},\zeta} e^{-i\omega_{\vec{k}}t} \varepsilon_{\vec{k},\zeta,\mu},\label{Eplus}
\ee
where $\zeta=1,2$\/ is the polarization index, $\mu=1,2,3$\/ is the Cartesian index, and 
$\varepsilon_{\vec{k},\zeta,\mu}$\/ is the (real) polarization vector \cite{GerryKnight2005}. For suitable broadband detectors,
the operator corresponding to the photoelectric current at time $t$\/ (in units of the electron charge) is
\be
\wh{I}(t)=\sum_{\nu,\mu}s_{\nu\mu}\;\wh{E}^{(-)}_\nu(t)\;\wh{E}^{(+)}_\mu(t)\label{Iopdef}
\ee
where
\be
\wh{E}^{(-)}_\mu(t)=\left(\wh{E}^{(+)}_\mu(t)\right)\da,\label{Eminus}
\ee
and where the sensitivity function $s_{\nu\mu}$\/ is constant and symmetric. 
Using (\ref{Eplus}) and (\ref{Eminus}) in (\ref{Iopdef}),
\be
\wh{I}(t)=\sum_{\vec{k},\zeta,\vec{l},\rho,\mu,\nu }\frac{\hbar}{2 \varepsilon_0 V}\;(\omega_{\vec{k}}\omega_{\vec{k}})^{1/2}
\;\wh{a}\da_{\vec{l},\rho}\wh{a}_{\vec{k},\zeta}\;e^{i(\omega_{\vec{l}}\;\;-\;\;\omega_{\vec{k}})t}\;
\varepsilon_{\vec{l},\rho,\nu}\varepsilon_{\vec{k},\zeta,\mu}s_{\nu \mu}.
\label{Iop}
\ee
The following functional of $\wh{I}(t)$\/ will play a key role (see Sec. \ref{Signal} below):
\be
\wh{S}=\tau^{-1}\int_0^\tau dt \; \cos(\omega_Ht+\theta_H)\wh{I}(t). \label{Sopdef}
\ee
Using (\ref{Iop}) in (\ref{Sopdef}),
\be
\wh{S}=\frac{\hbar}{4 \varepsilon_0 V}\sum_{\vec{k},\zeta,\vec{l},\rho,\mu,\nu
\mbox{ \footnotesize s.t. } |\omega_{\vec{l}}\;\;-\;\;\omega_{\vec{k}}|=\omega_H}
(\omega_{\vec{l}}\omega_{\vec{k}})^{1/2}\;\wh{a}\da_{\vec{l},\rho}\wh{a}_{\vec{k},\zeta}
\;e^{-i\varepsilon(\omega_{\vec{l}}\;\;-\;\;\omega_{\vec{k}})\theta_H}\; \varepsilon_{\vec{l},\rho\,\nu}\varepsilon_{\vec{k},\zeta\,\mu}s_{\nu\mu},\label{Sop}
\ee
where 
\be
\varepsilon(x)=\mbox{ sign of }x. \label{signfuncdef}
\ee
From (\ref{Sop}),
\bea
\wh{S}^2&=&\left(\frac{\hbar}{4 \varepsilon_0 V}\right)^2\sum_{\vec{k},\zeta,\vec{l},\rho,\mu,\nu
\mbox{ \footnotesize s.t. } |\omega_{\vec{l}}\;\;-\;\;\omega_{\vec{k}}|=\omega_H}
\;\;\sum_{\vec{k}\pr,\zeta\pr,\vec{l}\pr,\rho\pr,\mu\pr,\nu\pr
\mbox{ \footnotesize s.t. } |\omega_{\vec{l}\pr}\;\;-\;\;\omega_{\vec{k}\pr}|=\omega_H}
\;(\omega_{\vec{l}}\;\omega_{\vec{k}}\;\omega_{\vec{l}\pr}\;\omega_{\vec{k}\pr})^{1/2} \nonumber \\
\rule{0mm}{7mm}& & \wh{a}\da_{\vec{l},\rho}\wh{a}_{\vec{k},\zeta}\wh{a}\da_{\vec{l}\pr,\rho\pr}\wh{a}_{\vec{k}\pr,\zeta\pr}
\; e^{-i[\varepsilon(\omega_{\vec{l}}\;\;-\;\;\omega_{\vec{k}})+\varepsilon(\omega_{\vec{l}\pr}\;\;-\;\;\omega_{\vec{k}\pr})]\theta_H}\;
\varepsilon_{\vec{l},\rho,\nu}\varepsilon_{\vec{k},\zeta\,\mu}s_{\nu\mu}\;
\varepsilon_{\vec{l}\pr,\rho\pr,\nu\pr}\varepsilon_{\vec{k}\pr,\zeta\pr,\mu\pr}s_{\nu\pr\mu\pr}.\label{Ssqop}
\eea

\section{Heterodyne signal}\label{Signal}

As pointed out above, our motivation for investigating the use of squeezed light in
a heterodyne laser radar system is the presence of the essentially loss-free local oscillator beam.
In the heterodyne approach to target detection\cite{Kingston1978}, the mixing of the local oscillator with
the light reflected from the target generates a current in the photodetector which oscillates
with angular frequency $\omega_H$\/ and phase $\theta_H$\/, where $\omega_H$\/ and $\theta_H$\/
are respectively the frequency difference and phase difference (at the detector location)  between the light from the local
oscillator and the light reflected from the target.

Let $I(t)$\/ denote the total photoelectric current (in units of the electron charge) produced by the photodetector at time $t$\/ of a given
experimental run. Then the statistic $g$\/ we will use for heterodyne detection is the Fourier component
of $I(t)$\/ at a angular frequency $\omega_H$\/ and phase $\theta_H$\/, which we will denote by $S$:\/
\be
S=\tau^{-1}\int_0^\tau dt \; \cos(\omega_Ht+\theta_H)I(t). \label{Sdef}
\ee
The average of this signal  is
\be
\la S \ra= \tau^{-1} \int_0^\tau dt \; \cos(\omega_H t+\theta_H)\la I(t)\ra. \label{Savformula}
\ee
Here and subsequently, mean value is denoted by angle brackets ``$\la \; \ra$\/.'' 
We assume that $\omega_H$\/ and $\theta_H$\/ do  not vary from one run of the experiment to another; i.e.,
\bea 
\la \omega_H \ra &=&\omega_H,  \label{omegaHconst}\\
\la \theta_H \ra &=&\theta_H. \label{thetaHconst}
\eea

\section{Heterodyne target detection with squeezed local oscillator}\label{SecHettargdet}

Let $|0\ra_{\vec{k},\zeta}$\/ denote the vacuum state for mode $\vec{k},\zeta$\/ of the
electromagnetic field.  We consider the class of squeezed states \cite{GerryKnight2005} parameterized by
two complex numbers $\alpha$\/, $\xi$\/ and defined as
\be 
|\alpha,\xi\ra_{\vec{k},\zeta}=\wh{D}(\alpha)_{\vec{k},\zeta}\wh{Q}(\xi)_{\vec{k},\zeta}|0\ra_{\vec{k},\zeta}, \label{squeezedstatedef}
\ee
where the displacement operator $\wh{D}(\alpha)_{\vec{k},\zeta}$\/ and squeezing operator $\wh{Q}(\xi)_{\vec{k},\zeta}$\/
for mode $\vec{k},\zeta$\/ are, respectively,
\be
\wh{D}(\alpha)_{\vec{k},\zeta}=\exp\left(\alpha \wh{a}_{\vec{k},\zeta}\da-\alpha\as \wh{a}_{\vec{k},\zeta}\right),\label{displacementoperatordef}
\ee
\be
\wh{Q}(\xi)_{\vec{k},\zeta}=\exp\left[\frac{1}{2}\left(\xi\as\wh{a}_{\vec{k},\zeta}^2-\xi\left(\wh{a}_{\vec{k},\zeta}\da\right)^2\right)\right].
\label{squeezingoperatordef}
\ee
For $\xi=0$\/ the squeezed state $|\alpha,\xi\ra_{\vec{k},\zeta}$\/ reduces to the coherent state $|\alpha\ra_{\vec{k},\zeta}$\/:
\be
|\alpha\ra_{\vec{k},\zeta}=|\alpha,0\ra_{\vec{k},\zeta}=\wh{D}(\alpha)_{\vec{k},\zeta}|0\ra_{\vec{k},\zeta}.
\label{coherentstatedef}
\ee

So, in the absence of the coherent signal reflected from the target (null hypothesis, $H_0$\/), the full quantum state is
\be
|\psi_0\ra=|\alpha,\xi\ra _{LO} \prod_{\vec{k},\zeta\neq LO} |0\ra_{\vec{k},\zeta}.\label{psi0}
\ee
When the signal is present (alternative hypothesis, $H_1$\/), the state is
\be
|\psi_1\ra=|\beta\ra_T|\alpha,\xi\ra _{LO} \prod_{\vec{k},\zeta\neq T,LO} |0\ra_{\vec{k},\zeta}.\label{psi1}
\ee
Here $T$\/ and $LO$\/ are shorthand for  $\vec{k},\zeta$\/  for the target return signal
and local oscillator, respectively.

Taking for concreteness $\omega_T-\omega_{LO}=\omega_H > 0$\/, and using (\ref{Sop}), (\ref{psi0}) and
(\ref{psi1}),
\be
\la \psi_0|\wh{S}|\psi_0 \ra=0,\label{meanS0}
\ee
since the only possible nonzero term, $\wh{a}\da_{LO}\wh{a}_{LO}$\/, is forbidden by the restriction on
the summation  in (\ref{Sop}),
and
\be
\la \psi_1|\wh{S}|\psi_1 \ra=\frac{\kappa \hbar}{2\varepsilon_0 V}\;\left(\omega_T\omega_{LO}\right)^{1/2}\;
|\alpha||\beta|\cos(\theta_T-\theta_{LO}+\theta_H)\label{meanS1}
\ee
Assuming
\be
\omega_H \ll \omega_T,\hspace*{5mm} \omega_H \ll \omega_{LO},
\label{smallomegaH}
\ee
so
\be
\omega_T\approx\omega_{LO}\equiv\omega,\label{omega}
\ee
(\ref{meanS1}) becomes
\be
\la \psi_1|\wh{S}|\psi_1 \ra=\frac{\kappa \hbar\omega}{2\varepsilon_0 V}\;
|\alpha||\beta|\cos(\theta_T-\theta_{LO}+\theta_H)\label{meanS1b}
\ee
Here
\be
\theta_T=\arg{\beta},\hspace*{5mm}\theta_{LO}=\arg{\alpha},\label{thetadefs}
\ee
and
\be
\kappa=\sum_{\nu,\mu}\; \varepsilon_{LO,\nu}\;\varepsilon_{T,\mu}s_{\nu\mu}.\label{kappa}
\ee

The analysis presented here has left out phase factors in directions normal to the detector surface. Had those been
included,  the argument of the
the cosine in (\ref{meanS1}) would have terms dependent on the location on the surface of the
detector, unless both wave vectors, $T$\/and $LO$\/, were normal to the detector surface. So, unless
both wave vectors are sufficiently close to normal so that the respective wave fronts are parallel to the detector surface to within
less than a quarter wavelength over the surface, the net signal from the entire detector surface will add to zero  (i.e., zero
mixing efficiency). Assuming that
the polarization of the target return signal is also parallel to that of the local oscillator, we can rewrite (\ref{kappa})
as
\be
\kappa=\sum_{\nu,\mu}\; \varepsilon_{LO,\nu}\;\varepsilon_{LO,\mu}s_{\nu\mu}.\label{kappa2}
\ee

Using (\ref{Ssqop}) and (\ref{psi0}),
\bea
\la \psi_0|\wh{S}^2|\psi_0 \ra&=&\left(\frac{\hbar}{4 \varepsilon_0 V}\right)^2
\sum_{\vec{k},\zeta,\mu,\nu \mbox{ \footnotesize s.t. }|\omega_{LO}-\omega_{\vec{k}}|=\omega_H}\;\;
\sum_{\vec{l}\pr,\rho\pr,\mu\pr,\nu\pr \mbox{ \footnotesize s.t. }|\omega_{\vec{l}\pr}-\omega_{LO}|=\omega_H}\;\; 
\omega_{LO}\left(\omega_{\vec{k}}\omega_{\vec{l}\pr}\right)^{1/2}\nonumber\\
\rule{0mm}{7mm}&&\la\psi_0|\wh{a}\da_{LO}\wh{a}_{\vec{k},\zeta}\wh{a}\da_{\vec{l}\pr,\rho\pr}\wh{a}_{LO}|\psi_0\ra
\; e^{-i[\varepsilon(\omega_{LO}-\omega_{\vec{k}})+\varepsilon(\omega_{\vec{l}\pr}-\omega_{LO})]\theta_H}\nonumber\\
\rule{0mm}{7mm}&&\varepsilon_{LO,\nu}\varepsilon_{\vec{k},\zeta\,\mu}s_{\nu\mu}\;
\varepsilon_{\vec{l}\pr,\rho\pr,\nu\pr}\varepsilon_{LO,\mu\pr}s_{\nu\pr\mu\pr}.\label{meanSsqa}
\eea
Neither $\vec{k},\zeta$\/ nor $\vec{l}\pr,\rho$\/ can be $LO$\/, due to the restrictions in the summations arising
from the heterodyning.  If $\vec{k},\zeta\neq\vec{l}\pr,\rho$\/ then $\wh{a}_{\vec{k},\zeta}$\/ and $\wh{a}\da_{\vec{l}\pr,\rho\pr}$\/
can commute, yielding zero since the non-$LO$\/ modes are in the vacuum state. So the only surviving terms are
those for which $\vec{k},\zeta=\vec{l}\pr,\rho$\/:
\bea
\la \psi_0|\wh{S}^2|\psi_0 \ra&=&\left(\frac{\hbar}{4 \varepsilon_0 V}\right)^2
\sum_{\vec{k},\zeta,\mbox{ \footnotesize s.t. }|\omega_{LO}-\omega_{\vec{k}}|=\omega_H}\;\;\omega_{LO}\omega_{\vec{k}}\;\;
\bar{n}_{LO} \left(\sum_{\nu,\mu}\varepsilon_{LO,\nu}\varepsilon_{\vec{k},\zeta,\mu}s_{\mu\nu}\right)^2,\label{meanSsqb}
\eea
or, using (\ref{smallomegaH}) and (\ref{omega}),
\bea
\la \psi_0|\wh{S}^2|\psi_0 \ra&=&2\left(\frac{\hbar\omega_{LO}}{4 \varepsilon_0 V}\right)^2
\;\;
\bar{n}_{LO}\;\;\sum_{\vec{k},\zeta,\mbox{ \footnotesize s.t. }\omega_{\vec{k}}=\omega_{LO}} \left(\sum_{\nu,\mu}\varepsilon_{LO,\nu}\varepsilon_{\vec{k},\zeta,\mu}s_{\mu\nu}\right)^2,\label{meanSsqc}
\eea
where
\be
\bar{n}_{LO}=\mbox{}_{LO}\la \alpha,\xi|\wh{n}_{LO}|\alpha,\xi\ra_{LO},\label{nLObar}
\ee
\be
\wh{n}_{LO}=\wh{a}\da_{LO}\wh{a}_{LO}.\label{nLOop}
\ee

In going from (\ref{meanSsqb}) to (\ref{meanSsqc}) the expression for $\la \psi_0|\wh{S}^2|\psi_0 \ra$\/ picked up a factor
of ``2.'' This is as a result of contributions in the sum in (\ref{meanSsqb}) coming not only from modes with
$\vec{k}$\/ such that $\omega_{\vec{k}}=\omega_{LO}+\omega_H=\omega_T$\/, but also from modes with 
$\vec{k}$\/ such that $\omega_{\vec{k}}=\omega_{LO}-\omega_H$\/. The latter modes are termed ``image band'' modes \cite{Haus2000}.

The argument about mixing efficiency doesn't apply here to the sum in parentheses in (\ref{meanSsqb}) since the nonzero terms
come from raising and lowering operators corresponding to the same mode.  Define
\be
(\kappa\pr)^2=\sum_{\vec{k},\zeta,\mbox{ \footnotesize s.t. }\omega_{\vec{k}}=\omega_{LO}} \left(\sum_{\nu,\mu}\varepsilon_{LO,\nu}\varepsilon_{\vec{k},\zeta,\mu}
s_{\mu\nu}\right)^2.\label{kappaprime}
\ee
Since the sum over $\vec{\kappa}, \zeta$\/ contains the term $\vec{k},\zeta=LO$\/
\be
(\kappa\pr)^2 \ge (\kappa)^2,\label{kappaprgkappa}
\ee
where $\kappa$\/ is as given in (\ref{kappa2}). 
If the field of view of the detector is such that the only modes satisfying the constraint $\omega_{\vec{k}}=\omega_{LO}$\/
have $\vec{k}$\/ colinear with the wave vector of the local oscillator, then 
\be
(\kappa\pr)^2=\kappa^2.\label{kappaprime2}
\ee
Let $L_{tr}$\/ be the transverse size of the quantization volume (here taken to be equal to the size of the detector), and
let $\Omega$\/ be the solid angle of the detector's field of view. Then, since the transverse components of the wave
vector are quantized in units of $2\pi/L_{tr}$\/ and are assumed to be much smaller than the other component (in the direction
of the local oscillator wave vector),
\be
(\kappa\pr)^2\approx\kappa^2\cal{N}\label{kappaprime3}
\ee
where
\bea
\cal{N}&=&\sum_{\vec{k}\mbox{ \footnotesize s.t. }\omega_{\vec{k}}=\omega_{LO}}\nonumber\\
\rule{0mm}{7mm}&\approx&\mbox{max}(|\vec{k}|^2\Omega/((2\pi)/L_{tr})^2,1)\nonumber\\
\rule{0mm}{7mm}&=&\mbox{max}((L_{tr}/\lambda)^2\Omega,1),\label{sumformula}
\eea
with 
\be
\lambda=2\pi c/\omega
\ee
the wavelength of the local oscillator
and
\be
[x]=\mbox{\rm integral part of $x$\/.}\label{floordef}
\ee
E.g., for $\lambda=10^{-6}$\/m, $L_{tr}=10^{-5}$\/m, and $\Omega$=(1 mrad)$\mbox{}^2$\/, $\cal{N}$\/
equals unity.  Changing $L_{tr}$\/ to $10^{-3}$\/m and $\Omega$\/ to (10 mrad)$\mbox{}^2$\/ changes $\cal{N}$\/ to 100.
For the remainder of this paper we will  set
\be
{\cal N}=1,\label{calNeq1}
\ee
implying that (\ref{kappaprime2}) holds.

Using (\ref{kappaprime}) and (\ref{kappaprime2}) in (\ref{meanSsqc}),
\bea
\la \psi_0|\wh{S}^2|\psi_0 \ra&=&2\left(\frac{\kappa\hbar\omega_{LO}}{4 \varepsilon_0 V}\right)^2
\;\;
\bar{n}_{LO}.\label{meanSsqd}
\eea

Using (\ref{meanS0}), (\ref{meanS1}), 
and (\ref{meanSsqd})  in (\ref{DS}),
\bea
D_S^2&=&\frac{2|\alpha|^2|\beta|^2 \cos^2(\theta_T-\theta_{LO}+\theta_H)}{\bar{n}_{LO}}\nonumber\\
&=&2\left(1-\frac{\sinh^2(r)}{\bar{n}_{LO}}\right)\bar{n}_T \cos^2(\theta_T-\theta_{LO}+\theta_H)\label{DS2}
\eea
where 
\be
\bar{n}_T=\mbox{}_T\la \beta|\wh{n}_T|\beta\ra_T,\label{nT}
\ee
\be
\wh{n}_T=\wh{a}\da_T\wh{a}_T,\label{nTop}
\ee
and $r$\/ is the squeezing parameter, nonnegative by definition:
\be
r=|\xi|. \label{rdef}
\ee
These quantities and (\ref{nLObar}), (\ref{nLOop}) satisfy the relations \cite{GerryKnight2005}
\be
\bar{n}_{LO}=|\alpha|^2+\sinh^2(r),\label{nLOalphaconstraint}
\ee
\be
\bar{n}_T=|\beta|^2,\label{coherentnumber}
\ee
which have been used in obtaining (\ref{DS2}).
From 
(\ref{DS2}) it is clear that squeezing the local oscillator {\em decreases}\/ the signal-to-noise ratio.

As can be seen from (\ref{nLOalphaconstraint}), the value of $r$\/ is constrained by
\be
\sinh^2(r) \le \bar{n}_{LO}.\label{sinhsqrlen}
\ee
The case of equality in (\ref{sinhsqrlen}) is termed the  ``squeezed vacuum.''

\section{SNR in terms of conventional parameters}\label{SecSNRconvparams}

For a mode with expectation value of the number $\bar{n}$\/, the total energy is $E=\hbar\omega\bar{n}$\/. Hence the average energy 
density is $\rho_E=E/V=\hbar\omega\bar{n}/V$\/, the average energy flux is $\Phi=\rho_Ec=\hbar\omega\bar{n}c/V$\/, and the average power is
$P=\Phi_EA=\hbar\omega\bar{n}cA/V$\/, where $A=L_{tr}^2$\/ is the area of the quantization region (transverse to the wave vector of the mode
in question). Using these with (\ref{meanS1}),
\be
\la\psi_1|\wh{S}|\psi_1\ra^2=\frac{1}{c^2A^2}\left(\frac{\kappa}{2\varepsilon_0}\right)^2P_{LO}P_T\cos^2(\theta_T-\theta_{LO}+\theta_H)\left(1-\frac{\sinh^2(r)}{\bar{n}_{LO}}\right)
\label{meanSsqconv}
\ee
Let $L$\/ be the dimension of the quantization region parallel to the wave vector,
\be
L=V/A,\label{Ldef}
\ee
and, following \cite{Haus2000}, define the quantization time $T$\/ to be the time light would traverse this length,
\be
T=L/c=V/cA.\label{Tdef}
\ee
Taking the effective bandwidth $B$\/ to be \cite{Kingston1978}
\be
B=1/2T,\label{Bdef}
\ee
(\ref{meanSsqd}) becomes
\be
\la \psi_0|\wh{S}^2|\psi_0\ra=\left(\frac{\kappa}{4\varepsilon_0}\right)^2\frac{4 B \hbar \omega}{c^2A^2}P_{LO},
\label{meanSsqe}
\ee
 so
\be
D_S^2=\frac{P_T}{ \hbar\omega B}\cos^2(\theta_T-\theta_{LO}+\theta_H)\left(1-\frac{\sinh^2(r)}{\bar{n}_{LO}}\right).\label{DSsqconv}
\ee

\section{Quantum efficiency}\label{SecQuanteff}

Quantum efficiency is a measure of the extent to which photons are lost to
the detection process. This loss is   incorporated in the model by
allowing both the target-return beam $T$\/ and the local-oscillator beam $LO$\/ to pass through a beam splitter
before reaching the detector \cite{GardinerZoller2004}.  (Naively, one might think that
the quantum efficiency would involve the sensitivity function.  As seen in Sec. 6, this is not the case, 
since both the signal and the noise
scale with the sensitivity function.)

Let $\wh{a}_{\vec{k},\zeta}$\/ be the lowering operator at the
detector. This is related to the operator at the input port
to the beam splitter by
\be
\wh{a}_{\vec{k},\zeta}=t_{\vec{k},\zeta}\wh{a}_{(in)\vec{k},\zeta}+r_{\vec{k},\zeta}\wh{a}_{(vac)\vec{k},\zeta}, \label{BSops}
\ee
 where $\wh{a}_{(in)\vec{k},\zeta}$\/ and $\wh{a}_{(vac)\vec{k},\zeta}$\/ are respectively the operators
at the input and vacuum ports of the beam splitter, and $t_{\vec{k},\zeta}$\/ and $r_{\vec{k},\zeta}$ are
the (possibly wavelength or polarization dependent) c-number transmission and reflection coefficients.

Using (\ref{BSops}) in (\ref{Sop}),  we find the heterodyne signal operator to be
\bea
\wh{S}_B&=&\frac{\hbar}{4\varepsilon_0 V}\sum_{\vec{k},\zeta,\vec{l},\rho, \mu , \nu \mbox{ \footnotesize s.t. } |\omega_{\vec{l}}\;-\;\omega_{\vec{k}}|=\omega_H}
(\omega_{\vec{l}}\;\omega_{\vec{k}})^{1/2} \nonumber \\
&&(t^\ast_{\vec{l},\rho}\wh{a}\da_{(in)\vec{l},\rho}+r^\ast_{\vec{l},\rho}\wh{a}\da_{(vac)\vec{l},\rho})\;
(t_{\vec{k},\zeta}\wh{a}_{(in)\vec{k},\zeta}+r_{\vec{k},\zeta}\wh{a}_{(vac)\vec{k},\zeta}) \nonumber \\
&&e^{-i\varepsilon(\omega_{\vec{l}}\;-\;\omega_{\vec{l}})\theta_H}\; \varepsilon_{\vec{l},\rho,\nu} \;\varepsilon_{\vec{k},\zeta,\mu }\;s_{\nu 
\mu} .\label{SBops}
\eea
 
The null-hypothesis and alternative-hypothesis states are, respectively,
\be
|\psi_{B,0}\ra=|\psi_{(in),0}\ra|\psi_{(vac),0}\ra,\label{psiB0}
\ee
 \be
|\psi_{B,1}\ra=|\psi_{(in),1}\ra|\psi_{(vac),0}\ra,
\ee
 where
\bea
|\psi_{(vac),0}\ra&=&\prod_{\vec{k},\zeta}|0\ra_{(vac)},\label{psivac0}\\
|\psi_{(in),0}\ra&=&\prod_{\vec{k},\zeta \neq LO}|0\ra_{(in),\vec{k},\zeta} |\alpha,\xi\ra_{(in),LO},\label{psiin0}\\
|\psi_{(in),1}\ra&=&\prod_{\vec{k},\zeta \neq LO,T}|0\ra_{(in),\vec{k},\zeta} |\alpha,\xi\ra_{(in),LO}|\beta\ra_{(in),T}.\label{psiin1}
\eea
 
Using (\ref{SBops})-(\ref{psiin1}), defining
\be
\theta_B=\arg(t^\ast_{LO}t_T),\label{thetaBdef}
\ee
 and taking the polarizations of the local oscillator and target beam to be the same, we obtain
\be
\la \psi_{B,0}|\wh{S}_B | \psi_{B,0}\ra=0,\label{SB0expval}
\ee
 \be
\la \psi_{B,1}|\wh{S}_B | \psi_{B,1}\ra=|t_{LO}t_T|\left(\frac{\kappa \hbar \omega}{2 \varepsilon_0 V}\right)(\bar{n}-\sinh^2(r))^{1/2}\bar{n}_T^{1/2}\cos(\theta_T-\theta_{LO}+\theta_H+\theta_B),
\label{SB1expval}
\ee
 \be
\la \psi_{B,0}|\wh{S}_B^2 | \psi_{B,0}\ra=2\;|t_{LO}|^2 \left(\frac{\kappa \hbar \omega}{4\varepsilon_0 V}\right)\bar{n}_{LO}.\label{SB1sqexpval}
\ee
 From (\ref{SB0expval}) and (\ref{SB1sqexpval})
\bea
\mbox{Var}_0 S_B&=&\la \psi_{B,0}|\wh{S}_B^2 | \psi_{B,0}\ra-\la \psi_{B,0}|\wh{S}_B | \psi_{B,0}\ra^2\nonumber \\
&=&2\;|t_{LO}|^2 \left(\frac{\kappa \hbar \omega}{4\varepsilon_0 V}\right)\bar{n}_{LO}.\label{Var0SB}
\eea
 Using (\ref{SB0expval}), (\ref{SB1expval}), and (\ref{Var0SB}), the SNR is
\bea
D^2_{S_B}&=&(\la \psi_{B,1}|\wh{S}_B | \psi_{B,1}\ra-\la \psi_{B,0}|\wh{S}_B | \psi_{B,0}\ra^2/\mbox{Var}_0 S_B.\nonumber\\
&=&|t_T|^2\; 2\;\left(1-\frac{\sinh^2(r)}{\bar{n}_{LO}}\right)\;\bar{n}_T \; \cos^2(\theta_T-\theta_{LO}+\theta_H+\theta_B). \label{DSB}
\eea
 Comparing (\ref{DSB}) with  (\ref{DS2}), we see that the norm-squared of the transmission coefficient enters as the quantum efficiency $\eta$:
\be
\eta=|t_T|^2.\label{etadef}
\ee

\section{Direction-finding with a split detector }\label{SecDirfind}

Since  the SNR for direct detection of a light beam is improved by squeezing (see Appendix), even when
only  a single mode is excited,
one might expect that it is possible to improve the SNR of
directional measurements using a single-mode squeezed beam
and a  split detector. As pointed out  in \cite{Fabreetal2000}, this is 
not the case, and one must use a beam with at least two transverse
modes.

\subsection{Transverse modes}

\subsubsection{Expansions of operators}

However many modes are in non-vacuum states, if we are to examine a detector with a response which varies in the transverse direction
we should expand the electric field in transverse modes. (If nothing
else, we may find in some situations that the expansion in tranverse modes is unnecessary.)

We take the positive-frequency part of the electric-field operator to be
\be
\wh{E}^{(+)}_\mu (t,x)=i\sum_{\vec{k},\zeta,m}\left(\frac{\hbar\omega_{\vec{k}}}{2\varepsilon_0 V}\right)^{1/2}
\wh{a}\da_{\vec{k},\zeta,m}(x)\;e^{-i\omega_{\vec{k}}t}\;\;u_{\vec{k},\zeta,m}(x) \;\varepsilon_{\vec{k},\zeta,\mu}, \label{multimodeE}
\ee
 where $x$\/ is the transverse coordinate---we will only consider modes depending on a single transverse 
coordinate---and the transverse modes are indexed by $m$\/.  The current operator is
\be
\wh{I}(t,x)=\sum_{\nu\mu} \; \wti{s}_{\nu,\mu} \wh{E}^{(-)}_{\nu}(t,x)\wh{E}^{(+)}_{\mu}(t,x), \label{multimodeI}
\ee
where
\be
\wti{s}_{\nu,\mu}=s_{\nu,\mu}/W \label{stildedef}
\ee
 with $W$\/ the beam width at the detector, and
\be
\wh{E}^{(-)}_{\mu}(t,x)=\left(\wh{E}^{(+)}_{\mu}(t,x)\right)\da.\label{multimodeEnegfreqdef}
\ee
 Using (\ref{multimodeE}), (\ref{multimodeI}) and (\ref{multimodeEnegfreqdef}),
\bea
\wh{I}(t,x)&=&\sum_{\wh{l},\rho,n,\wh{k},\zeta,m,\nu,\mu}\frac{\hbar}{2\varepsilon_0 V}\left(\omega_{\vec{l}}\;\omega_{\vec{k}}\right)^{1/2}
\wh{a}\da_{\vec{l},\rho,n}\wh{a}_{\vec{k},\zeta,m}e^{i(\omega_{\vec{l}}\; - \; \omega_{\vec{k}})t}\nonumber \\
&&u^\ast_{\vec{l},\rho,n}(x)u_{\vec{k},\zeta,m}(x)\varepsilon_{\vec{l},\rho,\nu}\varepsilon_{\vec{k},\zeta,\mu}\wti{s}_{\nu\mu}.\label{multimodeI2}
\eea
 
\subsubsection{Mode expansion for split detector scenario}

The following approach is taken in \cite{Trepsetal2002}: ``Let us consider a beam of light with an electric field distribution
given by $E(x)$\/. We can build an orthonormal basis of the transverse plane $\{u_i\}$\/ such that $u_0=E(x)/||E(x)||$\/ is the first vector;
$u_1$\/ is a ``flipped'' mode, given by $-u_0(x)$\/ for $x< 0$\/ and $u_0(x)$\/
for $x >0$\/\ldots and the other modes are chosen in order to form a basis.'' They go on to conclude
that in computing the noise in a split detector, it is sufficient to consider only $u_0$\/ and $u_1$\/, ignoring
the higher transverse modes.

Since the beam width at the detector is $W$\/, 
\bea
u_0(x)&=&1, \hspace*{5mm}-W/2 \le x \le W/2 \nonumber\\
      &=&0  \hspace*{5mm}\mbox{ otherwise } \label{u0def}
\eea
 \bea
u_1(x)&=&-1, \hspace*{5mm} -W/2 \le x \le 0 \nonumber\\
      &=&1, \hspace*{5mm} 0 < x \le W/2 \nonumber\\
      &=&0  \hspace*{5mm}\mbox{ otherwise }\label{u1def}
\eea

\subsection{Direct-detection direction-finding}\label{Directdetectiondirectionfinding}

In direct detection the local oscillator beam, which induces oscillations in the photoelectric current when
mixed with the reflected light from that target, is absent.
Therefore we do not include the factor $\cos(\omega_Ht+\theta_H)$\/ in the  definition of the 
signal operator.

Using (\ref{signfuncdef}), the operator corresponding to direct detection of the direction of beam arrival with the split detector can be written as
\be
\wh{S}\pr_{sp}=\int  dx \; \varepsilon(x) \frac{1}{\tau}\int_0^{\tau} dt\;\wh{I}(t,x).\label{homosplitS}
\ee

Using (\ref{multimodeI2}) in (\ref{homosplitS}),
\be
\wh{S}\pr_{sp}=\sum_{\vec{l},\rho,n,\vec{k},\zeta,m,\nu,\mu \mbox{ \footnotesize s. t. }\omega_{\vec{l}}\; =\;\omega_{\vec{k}}}
\frac{\hbar\omega_{\vec{l}}}{2\varepsilon_0 V}\;\wh{a}\da_{\vec{l},\rho,n}\wh{a}_{\vec{k},\zeta,m}\int dx \varepsilon(x) u^\ast_{\vec{l},\rho,n}(x)
u_{\vec{k},\zeta,m}(x) \varepsilon_{\vec{l},\rho,\nu}\varepsilon_{\vec{k},\zeta,\mu}\wti{s}_{\nu\mu}\label{homosplitS2}
\ee
in the limit $\tau \rightarrow \infty$\/.

\subsubsection{Single nonvacuum transverse mode}

The ``alternative hypothesis'' state, corresponding to the single mode beam being
displaced by an amount $\delta$\/ in the $x$\/ direction (we will always
take $\delta >0$\/) is
\be
|\psi\pr_{sp,1}\ra=\prod_{\vec{l}\ppr,\rho\ppr,n\ppr\neq T} |0\ra_{\vec{l}\ppr,\rho\ppr,n\ppr}|\alpha,\xi\ra_T.\label{onemodeH1state}
\ee
The transverse mode function for the target-return mode is the even function $u_0(x)$\/ displaced in the positive-$x$\/
direction a distance $\delta$\/:
\be
u_{T-1}(x)=u_0(x-\delta).\label{uTdef}
\ee
The null-hypothesis state is the same, but with $\delta=0$\/:
\be
|\psi\pr_{sp,0}\ra=|\psi\pr_{sp,1}\ra_{\delta=0}.\label{onemodeH0state}
\ee

From (\ref{u0def}) and (\ref{uTdef}),
\be
\int dx \; \varepsilon(x) |u_{T-1}(x)|^2=2\delta. \label{uTsqint}
\ee

Using (\ref{homosplitS2})-(\ref{uTsqint}),
\be
\la\psi\pr_{sp,0}| \wh{S}\pr_{sp} |\psi\pr_{sp,0}\ra=0,\label{onemodeS0val}
\ee
and 
\be
\la\psi\pr_{sp,1}| \wh{S}\pr_{sp} |\psi\pr_{sp,1}\ra=2\delta\frac{\wti{\kappa}\hbar \omega_T}{2\varepsilon_0 V}\bar{n}_T,\label{onemodeS1val}
\ee
where
\be
\wti{\kappa}=\kappa/W.\label{kappatildedef}
\ee
Using (\ref{homosplitS2})-(\ref{uTsqint}), and keeping in mind that only the $T$\/ is in a nonvacuum state,
$$
\la\psi\pr_{sp,0}| \wh{S}^{\prime 2}_{sp} |\psi\pr_{sp,0}\ra=
\sum_{\vec{k},\zeta,m,\nu,\mu \mbox{ \footnotesize s. t. }\omega_{\vec{k}}\; = \; \omega_{T}}
\; \; 
\sum_{\vec{l}\pr,\rho\pr,n\pr,\nu\pr,\mu\pr \mbox{ \footnotesize s. t. }\omega_{\vec{l}\pr}\; = \; \omega_{T}} 
$$
$$
\prod_{\vec{l}\ppr,\rho\ppr,n\ppr \neq T}\mbox{}_{\vec{l}\ppr,\rho\ppr,n\ppr }\la 0|\;\;\mbox{}_T \la \alpha,\xi|
\;\;\wh{a}\da_T\wh{a}_{\vec{k}\zeta,m}\wh{a}\da_{\vec{l}\pr\rho\pr,n\pr}\wh{a}_T
\prod_{\vec{k}\ppr,\zeta\ppr,m\ppr \neq T}| 0\ra_{\vec{k}\ppr,\zeta\ppr,m\ppr } |\alpha,\xi\ra_T
$$
$$
\left(\int dx \; \varepsilon(x)u^\ast_T(x)u_{\vec{k},\zeta,m}(x)\right)
\left(\int dx\pr \; \varepsilon(x)u^\ast_{\vec{l}\pr,\rho\pr,n\pr}(x)u_T(x)\right)
$$
\be
\varepsilon_{T,\nu}\varepsilon_{\vec{k},\zeta,\mu}\wti{s}_{\nu\mu}
\;\;\varepsilon_{\vec{l}\pr,\rho\pr,\nu\pr}\varepsilon_{T,\mu\pr}
\wti{s}_{\nu\pr\mu\pr}
\ee
By virtue of the constraints in the summations as well as the $x,x\pr$\/ integrals (recall (\ref{onemodeH0state}) and (\ref{uTsqint}))
the only mode which can make a nonzero contribution is the odd transverse mode $u_1(x)$\/ with frequency $\omega_T$\/. Since this mode is
in the vacuum state, we obtain
\be
\la\psi\pr_{sp,0}| \wh{S}^{\prime 2}_{sp} |\psi\pr_{sp,0}\ra=W^2\left(\frac{\wti{\kappa}\hbar\omega_T}{2\varepsilon_0V}\right)^2{\bar{n}}_T,\label{onemodeSsqval}
\ee
Using (\ref{onemodeS0val}) and (\ref{onemodeSsqval}),
\bea
\mbox{Var}_0 S\pr_{sp\pr}&=&\la\psi\pr_{sp,0}| \wh{S}^{\prime 2}_{sp} |\psi\pr_{sp,0}\ra-\la\psi\pr_{sp,0}| \wh{S}^{\prime}_{sp} |\psi\pr_{sp,0}\ra^2
\nonumber\\
&=&W^2\left(\frac{\wti{\kappa}\hbar\omega_T}{2\varepsilon_0 V}\right)^2{\bar{n}}_T.\label{Varps0Sprsp}
\eea
Using (\ref{onemodeS0val}),(\ref{onemodeS1val}) and (\ref{Varps0Sprsp}), the SNR is
\be
D^2_{S\pr_{sp\pr}}=\left(\frac{2 \delta}{W}\right)^2 \bar{n}_T. \label{onemodeSNR}
\ee
In terms of the angular displacement $\Delta\theta$\/ and wavelength $\lambda$\/ of the beam and the focal length $f$\/
and aperture $d$\/ of the detector optics,
\be
\delta=f\Delta\theta,\label{deltaformula}
\ee
\be
W=f\lambda/d\label{Wformula}
\ee
(see, e.g., \cite{SmithThomson1988}), so (\ref{onemodeSNR}) can be written as
\be
D^2_{S\pr_{sp\pr}}=\left(\frac{2 d \Delta\theta}{\lambda}\right)^2 \bar{n}_T. \label{onemodeSNR2}
\ee
The minimum discernable angular beam displacement is defined as that angle for which the SNR=1. From (\ref{onemodeSNR2}),
\be
\Delta\theta_{\min\pr}=\frac{1}{2\;\bar{n}_T^{1/2}}\left(\frac{\lambda}{d}\right).\label{Deltathetamin}
\ee
Both (\ref{onemodeSNR})and  (\ref{Deltathetamin}) are seen to be unaffected by squeezing.

(Note that this analysis, as well as the ones that follow, use the ``top-hat'' form $u_0(x)$\/ for the
average beam profile.  A more realistic form for $u_0(x)$\/ and consequently for $u_1(x)$\/ would
likely change constant factors such as the ``1/2'' in (\ref{Deltathetamin}), but probably not the
dependence on occupation number or squeezing parameters.)

\subsubsection{Two nonvacuum transverse modes }\label{Twonontransverse}

We now examine a model along the lines of the experiment of \cite{Trepsetal2002}. That experiment uses a beam with
a ``flipped'' (i.e. odd function of $x$\/) coherent mode and an even-in-$x$\/ squeezed vacuum mode. The  beams are combined using a beam splitter
which reflects most of the squeezed vacuum and transmits a small part of the flipped coherent beam.
The coherent beam shifts in the transverse direction while the squeezed beam remains fixed, so we
will refer to the squeezed beam as the ``local oscillator''  and the flipped coherent beam as the ``target-return
beam.'' Both modes have the same wavelength. So, the alternative-hypothesis state is
\be
|\psi_{sp,1}\ra=\prod_{\vec{l}\ppr,\rho\ppr,n\ppr\neq LO, T} |0\ra_{\vec{l}\ppr,\rho\ppr,n\ppr}|\alpha,\xi\ra_{LO}|\beta\ra_T.\label{twomodeH1state}
\ee
The transverse mode functions are
\be
u\pr_{LO}(x)=u_0(x),\label{twomodeuLO}
\ee
\be
u\pr_T(x)=u_1(x-\delta).\label{twomodeuT}
\ee
As before, the null-hypothesis state is the same with $\delta=0$\/:
\be
|\psi_{sp,0}\ra=|\psi_{sp,1}\ra_{\delta=0}.\label{twomodeH0state}
\ee
Using (\ref{twomodeuLO}) and (\ref{twomodeuT}),
\be
\int dx \; \varepsilon(x){u\pr}^\ast_{LO}(x)u\pr_{LO}(x)=0,\label{uLOuLOint}
\ee
\be
\int dx \; \varepsilon(x){u\pr}^\ast_{LO}(x){u\pr}_{T}(x)=\int dx \; \varepsilon(x){u\pr}^\ast_{T}(x){u\pr}_{LO}(x)=W-3\delta,\label{uLOuTint}
\ee
\be
\int dx \; \varepsilon(x){u\pr}^\ast_{T}(x){u\pr}_{T}(x)=2\delta.\label{uTuTint}
\ee

Using (\ref{homosplitS}), (\ref{twomodeH1state}) and (\ref{twomodeH0state})-(\ref{uTuTint}),
\be
\la\psi_{sp,0}|\wh{S}\pr_{sp}|\psi_{sp,0}\ra=2W\left(\frac{\wti{\kappa}\hbar\omega}{2\varepsilon_0 V}\right)
(\bar{n}_{LO}-\sinh^2(r))^{1/2}\bar{n}^{1/2}\cos(\theta_T-\theta_{LO}),\label{twomodeSprspH0}
\ee
\be
\la\psi_{sp,1}|\wh{S}\pr_{sp}|\psi_{sp,1}\ra=\left(\frac{\wti{\kappa}\hbar\omega}{2\varepsilon_0 V}\right)
\left(2\delta\bar{n}_T+2(W-3\delta)(\bar{n}_{LO}-\sinh^2(r))^{1/2}\bar{n}^{1/2}\cos(\theta_T-\theta_{LO})\right),\label{twomodeSprspH1}
\ee
\bea
\mbox{Var}_0 S\pr_{sp}&=&\la\psi_{sp,1}|\wh{S}^{\prime 2}_{sp}|\psi_{sp,1}\ra-\la\psi_{sp,1}|\wh{S}^{\prime }_{sp}|\psi_{sp,1}\ra^2
\nonumber\\
&=&\left(\frac{\wti{\kappa}\hbar\omega}{2\varepsilon_0 V}\right)W^2\nonumber\\
&&\left( \bar{n}_{LO}+\bar{n}_T\left\{1+2\sinh(r)\left[\sinh(r)-\cosh(r)\cos(2\theta_T-\theta_{sq})\right]\right\}\right).\label{twomodeVarsp}
\eea
Choosing $\theta_{sq}$\/ so as to minimize (\ref{twomodeVarsp}), i.e., so that
\be
\cos(2\theta_T-\theta_{sq})=1,\label{twomodethetasqcond}
\ee
(\ref{twomodeVarsp}) becomes
\be
\mbox{Var}_0 S\pr_{sp}=\left(\frac{\wti{\kappa}\hbar\omega}{2\varepsilon_0 V}\right)W^2
\left( \bar{n}_{LO}+\bar{n}_Te^{-2r}\right).
\label{twomodeVarsp2}
\ee
Using (\ref{twomodeVarsp2}), (\ref{twomodeSprspH0}) and (\ref{twomodeSprspH1}), the SNR is
\be
D^2_{S\pr_{sp}}=\left(\frac{2\delta}{W}\right)^2\frac{\left(\bar{n}_T-3\left(\bar{n}_{LO}-\sinh^2(r)\right)^{1/2}\bar{n}_T^{1/2}
\cos(\theta_T-\theta_{LO})\right)^2}{\left( \rule[-2.5mm]{0cm}{.3cm}\bar{n}_{LO}+\bar{n}_Te^{-2r}\right)}.\label{D2Sprsp}
\ee

In this experiment the $LO$\/ mode is a squeezed vacuum,
\be
\bar{n}_{LO}=\sinh^2(r). \label{sqvccond}
\ee
For 
\be
\bar{n}_{LO} \gg 1, \label{largenLO}
\ee
(\ref{sqvccond}) implies
\be
e^{-2r}\approx\frac{1}{4\bar{n}_{LO}}.\label{largenLOrapprox}
\ee
Using (\ref{sqvccond})-(\ref{largenLOrapprox}) in (\ref{D2Sprsp})
\be
D^2_{S\pr_{sp}}\approx\left(\frac{2\delta}{W}\right)^2\bar{n}_T\left(\frac{\bar{n}_T}{\bar{n}_{LO}}\right).\label{D2Sprsplim}
\ee

This particular experiment isn't relevant to laser radar since it involves transmitting squeezed light,
but the variation below is.

\subsection{Heterodyne direction-finding} \label{SecHetDirFind}

To convert the model of the previous section into something that might be relevant for
laser radar, we need to insure that a) squeezed light is only used in the local oscillator, and b) the target-return
mode is not ``flipped,'' i.e., does not contain a sharp edge in the transverse direction where the phase of the
field changes abruptly. Such an edge would become diffuse over distance, due to diffraction;
and  it 
would be improbable to have it well-aligned with the boundary between the two halves of the split detector. 
In addition, c) the frequencies of $LO$\/ and $T$\/ must differ, so that they can be combined using a etalon rather than a beamsplitter.

(Cases  with smooth transition of the flipped mode are  analyzed in 
\cite{Fabreetal2000}.
These only yield factors of 0.60 and 0.94 in the minimum measurable distance
compared to coherent light. ``This modest improvement with respect to the standard quantum limit is due to the fact that the variation of the odd squeezed mode amplitude is too slow when one crosses the edge $x=0$\/.''\cite{Fabreetal2000})

According to \cite{Trepsetal2002},  switching which
mode, squeezed or coherent, is the flipped one makes no difference (in the homodyne situation considered in that reference; see previous section). 
So here we consider a squeezed local oscillator with an odd transverse mode function fixed so that the abrupt phase transition
always coincides with the split in the detector ($x=0$)\/. The target return beam is taken to be coherent with a flat transverse
profile, which is displaced from symmetry about $x=0$\/ by an amount $\delta > 0$\/.  That is,
\be
u_{LO}(x)=u_1(x),\label{uLOhet}
\ee
\be
u_{T}(x)=u_0(x-\delta),\label{uThet}
\ee

The signal operator is
\be
\wh{S}_{sp}=\int dx\;\varepsilon(x) \frac{1}{\tau}\int_0^{\tau}dt \cos(\omega_H t+\theta_H)\wh{I}(t).
\label{Ssp}
\ee
Using (\ref{multimodeI2}) in (\ref{Ssp})  
\bea
\wh{S}_{sp}&=&\sum_{\vec{l},\rho,n,\vec{k},\zeta,m,\nu,\mu \mbox{ \footnotesize s. t. } |\omega_{\vec{l}}\; - \;\omega_{\vec{k}}|=\omega_H}
\left(\frac{\hbar}{4\varepsilon_0 V}\right)\left(\omega_{\vec{l}}\;\omega_{\vec{k}}\right)^{1/2}\wh{a}\da_{\vec{l},\rho,n}\wh{a}_{\vec{k},\zeta,m}\nonumber\\
&&
e^{-i\varepsilon(\omega_{\vec{l}}\;-\;\omega_{\vec{k}})\theta_H}
\left(\int dx \; \varepsilon(x)u^\ast_{\vec{l},\rho,n}(x)u_{\vec{k},\zeta,m}(x)\right)
\varepsilon_{\vec{l},\rho,\nu}\varepsilon_{\vec{k},\zeta,\mu}s_{\nu\mu}.\label{Sspop}
\eea
The expressions for the states are the same  as in the homodyne case (eqs. (\ref{twomodeH1state}) and 
(\ref{twomodeH0state})), but with different transverse mode functions (\ref{uLOhet}) and (\ref{uThet}),
and of course different frequencies for the $LO$\/ and $T$\/ modes. 

An analysis along the lines of that in the previous section shows that
\be
\la\psi_{sp,0}|\wh{S}_{sp}|\psi_{sp,0}\ra=W\frac{\wti{\kappa}\hbar\omega}{2\varepsilon_0 V}\left(\bar{n}_{LO}-\sinh^2(r)\right)^{1/2}
\bar{n}_t^{1/2} \cos(\theta_t-\theta_{LO}+\theta_H),\label{hetSspH0}
\ee
\be
\la\psi_{sp,1}|\wh{S}_{sp}|\psi_{sp,1}\ra=\left(W-\delta\right) \frac{\wti{\kappa}\hbar\omega}{2\varepsilon_0 V}\left(\bar{n}_{LO}-\sinh^2(r)\right)^{1/2}
\bar{n}_t^{1/2} \cos(\theta_t-\theta_{LO}+\theta_H),\label{hetSspH1}
\ee
\bea
\mbox{Var}_0 S_{sp}&=&W^2\left(\frac{\wti{\kappa}\hbar\omega}{4\varepsilon_0 V}\right)^2
\left\{ \rule[-1mm]{0mm}{5mm} 2\bar{n}_{LO} \right.\nonumber\\
&&\left.+ 2\bar{n}_{T}\left[1-\sinh(r)\left(-\sinh(r)+\cosh(r)\cos(2\theta-\theta_{sq}+2\theta_H)\right)\right] \rule[-1mm]{0mm}{5mm}\right\}
\label{Var0Ssq}
\eea
Minimizing this by setting $\theta_{sq}$\/ so that
\be
2\theta-\theta_{sq}+2\theta_H=2\pi n, \hspace*{5mm} n=0,\pm 1, \pm2,\ldots \label{Var0Ssqmincond}
\ee
we obtain
\be
\mbox{Var}_0 S_{sp}=2W^2\left(\frac{\wti{\kappa}\hbar\omega}{4\varepsilon_0 V}\right)^2
\left[  \bar{n}_{LO} + \frac{\bar{n}_{T}}{2}\left(1+e^{-2r}\right) \right]
\label{Var0Ssq2}
\ee
The SNR is then
\bea
D^2_{S_{sp}}&=&\left(\la\psi_{sp,1}|\wh{S}_{sp}|\psi_{sp,1}\ra-\la\psi_{sp,0}|\wh{S}_{sp}|\psi_{sp,0}\ra\right)^2/\mbox{Var}_0 S_{sp}
\nonumber\\
&=&2\left(\frac{d \Delta\theta}{\lambda}\right)^2
\left(\frac{1-\frac{\sinh^2(r)}{\bar{n}_{LO}}}{1+\frac{\bar{n}_T}{2\bar{n}_{LO}}\left(1+e^{-2r}\right)}\right)
\bar{n}_T\cos^2\left(\theta_{T}-\theta_{LO}+\theta_H\right),\label{D2Ssp}
\eea
using (\ref{deltaformula}), (\ref{Wformula}), (\ref{hetSspH0}), (\ref{hetSspH1}) and (\ref{Var0Ssq2}).

The largest value of the 2nd factor in parentheses in (\ref{D2Ssp}) is unity, which can only be reached by
increasing $\bar{n}_{LO}$\/ so that
\be
\bar{n}_T/\bar{n}_{LO} \rightarrow 0\/,
\label{nbarTovernbarLOgoto0}
\ee
\be
\sinh^2(r)/\bar{n}_{LO} \rightarrow 0\/.
\label{sinhsqrovernbarLOgoto0}
\ee
The latter limit will be approached faster for {\em smaller}\/ $r$\/.  There is presumably no practical
limit to how large the local-oscillator signal can be made relative to the target-return signal in a laser radar system. So, there is no
reason to do squeezing.

Essentially the same result is obtained taking the transverse functions for both $LO$\/ and $T$\/ to be even ($u_0(x)$\/):
\be
D^2_{S_{sp-2e}}=2\left(\frac{d \Delta\theta}{\lambda}\right)^2
\left(\frac{1-\frac{\sinh^2(r)}{\bar{n}_{LO}}}{1+\frac{\bar{n}_T}{\bar{n}_{LO}}\left(1+e^{-2r}\right)}\right)
\bar{n}_T\cos^2\left(\theta_{T}-\theta_{LO}+\theta_H\right).\label{D2Sspeven}
\ee

\section{Heterodyne phase change estimation}\label{SecHetphase}

Consider again the heterodyne target scenario of Sec. \ref{SecHettargdet}, and take the null-hypothesis state to
be the alternative-hypothesis state of that section:
\be
|\psi_{0,ph}\ra=|\beta\ra_T|\alpha,\xi\ra _{LO} \prod_{\vec{k},\zeta\neq T,LO} |0\ra_{\vec{k},\zeta}.\label{psi0ph}
\ee
Take the alternative-hypothesis state to be
\be
|\psi_{1,ph}\ra=|\beta e^{i\delta\theta_T}\ra_T|\alpha,\xi\ra _{LO} \prod_{\vec{k},\zeta\neq T,LO} |0\ra_{\vec{k},\zeta}.\label{psi1ph}
\ee
I. e., the null hypothesis is that there is a target present at a distance such that the phase of
the target-return signal at the detector is $\arg \beta$\/; the alternative hypothesis is identical to the
null hypothesis, but with a value of the phase differing by an amount $\delta\theta_T$\/.

The SNR for discrimination between these two hypotheses---i.e., for detecting a small change in
the phase of the target-return signal---is
\be
D^2_{ph}=\left(\la\psi_{1,ph}|  \wh{S}|\psi_{1,ph}\ra  -  \la\psi_{0,ph}|\wh{S}  |\psi_{0,ph}\ra \right)^2/\mbox{Var}_{0,ph} S,\label{D2Sphdef}
\ee
where
\be
\mbox{Var}_{0,ph} S=\la\psi_{1,ph}|  \wh{S}^2|\psi_{1,ph}\ra - \la\psi_{1,ph}|  \wh{S}|\psi_{1,ph}\ra^2.\label{Var0Sphdef}
\ee
Using (\ref{Sop}), (\ref{psi1}), (\ref{meanS1b}), (\ref{psi0ph}) and (\ref{psi1ph}),\vspace*{3mm}
$$
\hspace*{-3in}\la\psi_{1,ph}|  \wh{S}|\psi_{1,ph}\ra  -  \la\psi_{0,ph}|\wh{S}  |\psi_{0,ph}\ra 
$$
\bea
&=&\frac{\kappa \hbar\omega}{2\varepsilon_0 V}\;
|\alpha||\beta|\left(\cos(\theta_T+\delta\theta_T-\theta_{LO}+\theta_H)-\cos(\theta_T-\theta_{LO}+\theta_H)\right)\nonumber\\
&\approx&-\delta\theta_T\frac{\kappa \hbar\omega}{2\varepsilon_0 V}\;
|\alpha||\beta|\sin(\theta_T-\theta_{LO}+\theta_H).\label{Var0Sphnumerator}
\eea
for small $\delta\theta_T$\/. Using (\ref{Ssqop}), (\ref{psi0ph}) 
and (\ref{Var0Sphdef}),
\bea
\mbox{Var}_{0,ph} S&=&2\left(\frac{\kappa\hbar\omega}{4\varepsilon_0 V}\right)^2\nonumber\\
&&\left( \bar{n}_{LO}+\bar{n}_T\left\{1+\sinh(r)\left[\sinh(r)-\cosh(r)\cos(2\theta_T-\theta_{sq}+2\theta_H)\right]\right\}\right)\label{Var0phgen}
\eea
Minimizing this by taking $\theta_{sq}$\/ to satisfy
\be
2\theta_T-\theta_{sq}+\theta_H=2\pi n, \hspace*{5mm} n=0,\pm 1,\pm2,\ldots\label{minimizeVar0sph}
\ee
we obtain
\be
\mbox{Var}_{0,ph} S=2\left(\frac{\kappa\hbar\omega}{4\varepsilon_0 V}\right)^2
\left[ \bar{n}_{LO}+\frac{\bar{n}_T}{2}\left(1+e^{-2r}\right)\right]\label{Var0phmin}
\ee
Using (\ref{D2Sphdef}), (\ref{Var0Sphnumerator}) and (\ref{Var0phmin}),
\be
D^2_{ph}=2\left(\delta\theta_T\right)^2
\left(\frac{1-\frac{\sinh^2(r)}{\bar{n}_{LO}}}{1+\frac{\bar{n}_T}{2\bar{n}_{LO}}\left(1+e^{-2r}\right)}\right)
\bar{n}_T\sin^2\left(\theta_{T}-\theta_{LO}+\theta_H\right)\label{D2phfinal}
\ee

As per the discussion at the end of Sec. \ref{SecHetDirFind}, there is nothing to be gained by squeezing the $LO$\/.

\section{Balanced detection}\label{SecBaldet}

In balanced detection \cite{YuenChan1983,AnnovazziLodietal1992}, the $LO$\/ and $T$\/ beams each enter through one of the input ports of a 50/50 beamsplitter,
and the signal is the difference of the photoelectron currents at detectors at the two output ports.

Operators corresponding to modes of the beams entering the input ports of the beamsplitter will be denoted by subscripts ``$(LO)$\/''
and ``$(T)$\/.'' Operators corresponding to modes of beams leaving the exit ports will be denoted by subscripts
``$tr$\/'' and ``$ref$\/'' (respectively ``transmitted'' and ``reflected'' relative to the target-return beam). Then
\be
\wh{a}_{(ref),\vec{k},\zeta}=r \wh{a}_{(T),\vec{k},\zeta}+t\pr\wh{a}_{(LO),\vec{k},\zeta}\label{arefdef}
\ee
\be
\wh{a}_{(tr),\vec{k},\zeta}=t \wh{a}_{(T),\vec{k},\zeta}+r\pr\wh{a}_{(LO),\vec{k},\zeta}\label{atrdef}
\ee
Using (\ref{arefdef}) and (\ref{atrdef}) in (\ref{Iop}), the current operators at the two photodetectors are, respectively,
$$
\wh{I}_{(tr)}(t)=\sum_{\vec{l}\rho,\vec{k}\zeta,\nu\mu}\frac{\hbar}{2\varepsilon_0 V}\left(\omega_{\vec{l}}\;\omega_{\vec{l}}\right)^{1/2}
e^{i(\omega_{\vec{l}}\;-\;\omega_{\vec{k}})t}\varepsilon_{\vec{l},\rho,\nu}\varepsilon_{\vec{k},\zeta,\mu}s_{\nu\mu}
$$
\be
\left(|t|^2{\wh{a}\da}_{(T)\vec{l},\rho}{\wh{a}}_{(T)\vec{k},\zeta}+r^{\prime\ast}t\;{\wh{a}\da}_{(LO)\vec{l}\rho}{\wh{a}}_{(T)\vec{k}\zeta}
+t^{\ast}r\pr\;{\wh{a}\da}_{(T)\vec{l}\rho}{\wh{a}}_{(LO)\vec{k}\zeta}+|r\pr|^2{\wh{a}\da}_{(LO)\vec{l},\rho}{\wh{a}}_{(LO)\vec{k},\zeta}\right),\label{Itrop}
\ee
$$
\wh{I}_{(ref)}(t)=\sum_{\vec{l}\rho,\vec{k}\zeta,\nu\mu}\frac{\hbar}{2\varepsilon_0 V}\left(\omega_{\vec{l}}\;\omega_{\vec{l}}\right)^{1/2}
e^{i(\omega_{\vec{l}}\;-\;\omega_{\vec{k}})t}\varepsilon_{\vec{l},\rho,\nu}\varepsilon_{\vec{k},\zeta,\mu}s_{\nu\mu}
$$
\be
\left(|r|^2{\wh{a}\da}_{(T)\vec{l},\rho}{\wh{a}}_{(T)\vec{k},\zeta}+t^{\prime\ast}r\;{\wh{a}\da}_{(LO)\vec{l}\rho}{\wh{a}}_{(T)\vec{k}\zeta}
+r^{\ast}t\pr\;{\wh{a}\da}_{(T)\vec{l}\rho}{\wh{a}}_{(LO)\vec{k}\zeta}+|t\pr|^2{\wh{a}\da}_{(LO)\vec{l},\rho}{\wh{a}}_{(LO)\vec{k},\zeta}\right).\label{Irefop}
\ee
The operator corresponding to the difference between the currents at the two detectors is 
\be
\wh{I}_{(diff)}(t)=\wh{I}_{(tr)}(t)-\wh{I}_{(ref)}(t).\label{Idiffdef}
\ee
From (\ref{Itrop})-(\ref{Idiffdef}),
$$
\wh{I}_{(diff)}(t)=
\sum_{\vec{l}\rho,\vec{k}\zeta,\nu\mu}\frac{\hbar}{2\varepsilon_0 V}\left(\omega_{\vec{l}}\;\omega_{\vec{l}}\right)^{1/2}
e^{i(\omega_{\vec{l}}\;-\;\omega_{\vec{k}})t}\varepsilon_{\vec{l},\rho,\nu}\varepsilon_{\vec{k},\zeta,\mu}s_{\nu\mu}
$$
$$
\left(\left(|t|^2-|r|^2\right){\wh{a}\da}_{(T)\vec{l},\rho}{\wh{a}}_{(T)\vec{k},\zeta}
+\left(\rule[-1mm]{0mm}{5mm}r^{\prime\ast}t-t^{\prime\ast}r\right)\;{\wh{a}\da}_{(LO)\vec{l}\rho}{\wh{a}}_{(T)\vec{k}\zeta}\right.
$$
\be
\left. \left(\rule[-1mm]{0mm}{5mm}t^{\ast}r\pr -r^{\ast}t\pr\right)\;{\wh{a}\da}_{(T)\vec{l}\rho}{\wh{a}}_{(LO)\vec{k}\zeta}
+\left(|r\pr|^2-|t\pr|^2\right){\wh{a}\da}_{(LO)\vec{l},\rho}{\wh{a}}_{(LO)\vec{k},\zeta}\right),\label{Idifop}
\ee
which simplifies to 
$$
\wh{I}_{(bal)}(t)=
-i\sum_{\vec{l}\rho,\vec{k}\zeta,\nu\mu}\frac{\hbar}{2\varepsilon_0 V}\left(\omega_{\vec{l}}\;\omega_{\vec{k}}\right)^{1/2}
e^{i(\omega_{\vec{l}}\;-\;\omega_{\vec{k}})t}\varepsilon_{\vec{l},\rho,\nu}\varepsilon_{\vec{k},\zeta,\mu}s_{\nu\mu}
$$
\be
\left({\wh{a}\da}_{(LO)\vec{l}\rho}{\wh{a}}_{(T)\vec{k}\zeta}
-{\wh{a}\da}_{(T)\vec{l}\rho}{\wh{a}}_{(LO)\vec{k}\zeta}\right)\label{Idifop2}
\ee
when the transmission and reflection coefficients are chosen appropriate to the balanced case; specifically,
\be
t=t\pr=1/\sqrt{2},\label{balcoeffst}
\ee
\be
r=r\pr=i/\sqrt{2}.\label{balcoeffsr}
\ee
The heterodyne signal operator is
\bea
\wh{S}_{bal-het}&=&\frac{1}{\tau}\int_0^{\tau}dt\; \cos(\omega_H+\theta_H)\wh{I}_{(bal)}(t)\nonumber\\
&=&-i\sum_{\vec{l}\rho,\vec{k}\zeta,\nu\mu \mbox{ \footnotesize s. t. } |\omega_{\vec{l}}\;-\;\omega_{\vec{k}}|=\omega_H}\frac{\hbar}{4\varepsilon_0 V}\left(\omega_{\vec{l}}\;\omega_{\vec{k}}\right)^{1/2}
e^{-i\varepsilon(\omega_{\vec{l}}\;-\;\omega_{\vec{k}})\theta_H}\varepsilon_{\vec{l},\rho,\nu}\varepsilon_{\vec{k},\zeta,\mu}s_{\nu\mu}\nonumber\\
&&\left({\wh{a}\da}_{(LO)\vec{l},\rho}{\wh{a}}_{(T)\vec{k},\zeta}
-{\wh{a}\da}_{(T)\vec{l},\rho}{\wh{a}}_{(LO)\vec{k},\zeta}\right).\label{hetbalSop}
\eea
The null-hypothesis and alternative-hypothesis states are, respectively,
\be
|\psi_{bal,0}\ra=\left(\prod _{\vec{l}\ppr,\rho\ppr}|0\ra_{(T)\vec{l}\ppr,\rho\ppr}\right)
                \left(\prod _{\vec{k}\ppr,\zeta\ppr \neq LO}|0\ra_{(LO)\vec{k}\ppr,\zeta\ppr}\right)
                |\alpha,\beta\ra_{(LO)LO},\label{psibal0}
\ee
\be
|\psi_{bal,1}\ra=\left(\prod _{\vec{l}\ppr,\rho\ppr \neq T}|0\ra_{(T)\vec{l}\ppr,\rho\ppr}\right)
                \left(\prod _{\vec{k}\ppr,\zeta\ppr \neq LO}|0\ra_{(LO)\vec{k}\ppr,\zeta\ppr}\right)
                |\alpha,\beta\ra_{(LO)LO}|\beta\ra_{(T)T},\label{psibal1}
\ee
Using (\ref{hetbalSop})-(\ref{psibal1}) the SNR is found to be
\be
D^2_{bal-het}=2\left(1-\frac{\sinh^2(r)}{\bar{n}_{LO}}\right)\bar{n}_T\sin^2(\theta_T-\theta_{LO}+\theta_H).\label{hetbalSNR}
\ee

For homodyne detection, the signal operator is
\bea
\wh{S}_{bal-hom}&=&\frac{1}{\tau}\int_0^{\tau}dt\; \wh{I}_{(bal)}(t)\nonumber\\
&=&-i\sum_{\vec{l}\rho,\vec{k}\zeta,\nu\mu \mbox{ \footnotesize s. t. } \omega_{\vec{l}}\;=\;\omega_{\vec{k}}}\frac{\hbar\omega_{\vec{l}}}{2\varepsilon_0 V}
\;\varepsilon_{\vec{l},\rho,\nu}\varepsilon_{\vec{k},\zeta,\mu}s_{\nu\mu}\nonumber\\
&&\left({\wh{a}\da}_{(LO)\vec{l},\rho}{\wh{a}}_{(T)\vec{k},\zeta}
-{\wh{a}\da}_{(T)\vec{l},\rho}{\wh{a}}_{(LO)\vec{k},\zeta}\right).\label{hombalSop}
\eea
The states $|\psi_{bal,0}\ra$\/ and $|\psi_{bal,1}\ra$\/ are the same as in the heterodyne case, with of the course $\omega_H=\omega_T$\/.
The resulting SNR is
\be
D^2_{bal-hom}=4\left(1-\frac{\sinh^2(r)}{\bar{n}_{LO}}\right)
\bar{n}_T\sin^2\left(\theta_{T}-\theta_{LO}\right).\label{hombalSNR}
\ee
Neither (\ref{hetbalSNR}) nor (\ref{hombalSNR}) is improved by squeezing.

\renewcommand{\theequation}{A-\arabic{equation}}
  \setcounter{equation}{0}  

\section*{Appendix: Direct target detection}\label{Appendix}
 
\subsection*{A.1. Signal-to-noise ratio}\label{Secdirectdet}

The states for  $H_0$\/ and $H_1$\/ are, respectively,
\be
|\psi_0\pr\ra= \prod_{\vec{k},\zeta} |0\ra_{\vec{k},\zeta},\label{psi0pr}
\ee
 and
\be
|\psi_1\pr\ra=|\alpha,\xi\ra _{T} \prod_{\vec{k},\zeta\neq T} |0\ra_{\vec{k},\zeta}.\label{psi1pr}
\ee
 The signal operator is 
\bea
\wh{S}\pr&=&\tau^{-1}\int_0^\tau\;dt\; \wh{I}(t)\nonumber\\
&=&\sum_{\vec{l},\rho,\vec{k},\zeta,\mu,\nu\mbox{ \footnotesize s.t. }\omega_{\vec{l}}\;=\;\omega_{\vec{k}}}
\frac{\hbar\omega_{\vec{k}}}{2 \varepsilon_0 V}\;\wh{a}\da_{\vec{l},\rho}\wh{a}_{\vec{k},\zeta}\;\varepsilon_{\vec{l},\rho\nu}\varepsilon_{\vec{k},\zeta\mu}s_{\nu\mu}.\label{Sprop}
\eea
(See comments at the beginning of Sec. \ref{Directdetectiondirectionfinding}.)
 Since the variance of the signal (\ref{Sprop}) vanishes in the state (\ref{psi0pr}), the definition of the equivalent signal-to-noise ratio must
be  changed to 
\be
{D\pr}_g^2=\frac{[E(g|H_1)-E(g|H_0)]^2}{\mbox{Var}_1g}\label{Dprgsq}
\ee
 with 
\be
\mbox{Var}_1g=E(g^2|H_1)-[E(g|H_1)]^2.\label{var1pr}
\ee
 This a reasonable definition, since variance in the signal under either hypothesis will contribute to detection errors.
Taking $g=S\pr$\/ and using (\ref{kappaprime2}) and (\ref{psi0pr})-(\ref{Dprgsq}),
\be
{D\pr}_{S\pr}^2=\frac{\bar{n}_T^2}{ \mbox{var}_{sq}(n_T) },\label{Dprgsqb}
\ee
 where $\mbox{var}_{sq}(n_T)$\/ is the variance of $\wh{n}_T$\/ in the squeezed state,
\be
\mbox{var}_{sq}(n_T)=\mbox{}_T\la\alpha,\xi|\wh{n}_T^2|\alpha,\xi\ra_T-\left(\mbox{}_T\la\alpha,\xi|\wh{n}_T|\alpha,\xi\ra_T\right)^2.\label{var1def}
\ee
 
For suitable choice of parameters of the squeezed state, $\mbox{var}_{sq}(n_T) <  \bar{n}_T$\/. So
squeezing can improve direct-detection SNR. For $|\alpha,\xi\ra_T$\/,
\be
\mbox{var}_{sq}(n_T)=(\bar{n}_T-\sinh^2(r))|\cosh(r)-e^{i(\theta_{sq}-2\theta_T)}\sinh(r)|^2+2\cosh^2(r)\sinh^2(r)\label{varsqn}
\ee
 (see, e.g., \cite{GerryKnight2005}), where $\theta_T$\/ is as given in (\ref{thetadefs}) and $\theta_{sq}$\/ is defined by
\be
\xi=re^{i\theta_{sq}}.\label{thetasqdef}
\ee
 To minimize (\ref{varsqn}) set $\theta_{sq}$\/ so that
\be
\theta_{sq}-2\theta_T=2\pi n,\hspace{5mm}n=0,\pm 1, \pm 2, \ldots ,\label{thetasqchoice}
\ee
 so
\be
\mbox{var}_{sq}(n_T)=(\bar{n}_T-\sinh^2(r))e^{-2r}+2\cosh^2(r)\sinh^2(r)\label{varsqn2}
\ee
 and
\be
{D\pr}_{S\pr}^2=\frac{\bar{n}_T^2}{(\bar{n}_T-\sinh^2(r))e^{-2r}+2\cosh^2(r)\sinh^2(r)}.\label{DprSpr}
\ee

Suppose the amount of squeezing, as quantified by the values of $r$\/,
is much less than the maximum allowed:
\be
\sinh^2(r) \ll \bar{n}_T.\label{sinhsqrlln}
\ee
Then (\ref{DprSpr}) becomes
\be
{D\pr}_{S\pr}^2=\bar{n}_T e^{2r}.\label{DprSprsinhsqrlln}
\ee
For large $\bar{n}_T$ and $r$\/,
\be
\sinh^2(r) \approx \cosh^2(r) \approx \frac{1}{4} e^{2r}, \label{sinhcoshlimit}
\ee
 so, with (\ref{sinhsqrlln}),
\be
e^{2r} \ll 4\bar{n}_T.\label{e2rlim}
\ee
 An upper bound on the SNR is therefore, using (\ref{DprSprsinhsqrlln}) and (\ref{e2rlim}),
\be
{D\pr}_{S\pr}^2 \ll 4\bar{n}_T^2.\label{DprSprupperlim}
\ee

\subsection*{A.2. Quantum efficiency in direct detection}
\label{Secquanteffdirectdet}
 
Using (\ref{BSops}) in (\ref{Sprop}), the direct-detection signal operator is
\bea
\wh{S}\pr_B&=&\sum_{\vec{l},\rho,\vec{k},\zeta, \mu, \nu \mbox{ \footnotesize s.t. }\omega_{\vec{l}}\;=\;\omega_{\vec{k}}}\left(\frac{\hbar \omega_{\vec{k}}}{2\varepsilon_0 V}\right)\nonumber\\
&&(t^\ast_{\vec{l},\rho}\wh{a}\da_{(in)\vec{l},\rho}+r^\ast_{\vec{l},\rho}\wh{a}\da_{(vac)\vec{l},\rho})\;
(t_{\vec{k},\zeta}\wh{a}_{(in)\vec{k},\zeta}+r_{\vec{k},\zeta}\wh{a}_{(vac)\vec{k},\zeta}) 
\varepsilon_{\vec{l},\rho,\nu} \;\varepsilon_{\vec{k},\zeta,\mu }\;s_{\nu 
\mu} \label{SBprop}
\eea

The  states are
\be
|\psi\pr_{B,0}\ra=|\psi\pr_{(in),0}\ra|\psi_{(vac),0}\ra,\label{psiprB0}
\ee
\be
|\psi\pr_{B,1}\ra=|\psi\pr_{(in),1}\ra|\psi_{(vac),0}\ra,
\ee
 where
\bea
|\psi\pr_{(in),0}\ra&=&\prod_{\vec{k},\zeta }|0\ra_{(in),\vec{k},\zeta} ,\label{psiprin0}\\
|\psi\pr_{(in),1}\ra&=&\prod_{\vec{k},\zeta \neq T}|0\ra_{(in),\vec{k},\zeta} \; |\alpha,\xi\ra_{(in),T}\label{psiprin1}
\eea
 and with $|\psi_{(vac),0}\ra$\/ as given in (\ref{psivac0}).

Using (\ref{Dprgsq}), (\ref{var1pr}), and (\ref{SBprop})-(\ref{psiprin1}),
\be
{D\pr}^2_{S\pr_B}=\frac{|t_T|^2\; \bar{n}_{(in),T}^2}{|t_T|^2\;\mbox{var}_{sq}n_{(in),T}+(1-|t_T|^2)\;\bar{n}_{(in),T}}\label{DprSprB}
\ee
 where $\mbox{var}_{sq}n_{(in),T}$\/ is given by (\ref{var1def}) with $\wh{n}_T \rightarrow \wh{n}_{(in),T}$\/.  Comparing with (\ref{Dprgsqb}) we see that nonunity quantum efficiency,  $\eta=|t_T|^2< 1,$\/ shifts the statistics of the noise
towards those of coherent-state light
(variance=$n_{(in),T}$\/).

\subsection*{A.3. Loss limits squeezed-light improvement in direct detection SNR}

The analysis of Sec. A.1 
shows that the use of squeezed light can improve direct-detection SNR compared to that obtained with coherent light.  From Sec. A.2 
above we
can see that the presence of loss places a limit on the amount of improvement which is possible.  

For coherent light, 
\be
\mbox{\rm var}_{sq}n_{(in),T}\rightarrow
\mbox{\rm var}_{coh}n_{(in),T}=\bar{n}_{(in),T},  \label{varcoh}
\ee
 so, from (\ref{DprSprB}),
\be
{D\pr}^2_{S\pr_B}\rightarrow{D\pr}^2_{S\pr_B-coh}=|t_T|^2\; \bar{n}_{(in),T}.
\label{Dprsqcoh}
\ee
 So the improvement in SNR obtained by using squeezed light is
\be
\frac{{D\pr}^2_{S\pr_B}}{{D\pr}^2_{S\pr_B-coh}}=
\frac{ \bar{n}_{(in),T}}{|t_T|^2\;\mbox{var}_{sq}n_{(in),T}+(1-|t_T|^2)\;\bar{n}_{(in),T}}\le \frac{ 1}{1-|t_T|^2}\label{dirimplim}
\ee
 since $\mbox{var}_{sq}n_{(in),T} \ge 0$\/, or
\be
\frac{{D\pr}^2_{S\pr_B}}{{D\pr}^2_{S\pr_B-coh}}\le \frac{1}{L} \label{dirimplim2}
\ee
 where the loss $L$\/ is 
\be
L=1-|t_T|^2=1-\eta\label{Ldef}
\ee
 by (\ref{etadef}).

\section*{Acknowledgments}

M. A. R. thanks Jonathan Ashcom and  Jae Kyung for a helpful discussion on mixing efficiency.
This work was sponsored by the Air Force under Air Force Contract 
FA8721-05-C-0002.  Opinions, interpretations, conclusions, and 
recommendations are those of the author and are not necessarily endorsed 
by the U.S. Government.


\begin{thebibliography}{99}
\bibitem{Caves1981}C.~M.~Caves,``Quantum-mechanical noise in an interferometer,'' Phys. Rev. D {\bf 23}, 1693-1708 (1981).
\bibitem{Kingston1978} R.~H.~Kingston, {\em Detection of Optical and Infrared Radiation}\/, (Springer 1978).
\bibitem{Lietal1997} Y.-q.~Li, P.~Lynam, M.~Xiao, and P.~J.~ Edwards, ``Sub-shot-noise laser Doppler anemometry with amplitude-squeezed light,'' Phys. Rev. Lett. {\bf 78}, 3105-3108 (1997).
\bibitem{Lietal1999}Y.-q.~Li, D.~Guzun,  and M.~Xiao, ``Sub-shot-noise-limited optical heterodyne detection using an amplitude-squeezed local oscillator,''
Phys. Rev. Lett.  {\bf 82}, 5225-5228 (1999).
\bibitem{RubinKaushik2007}M.~A.~Rubin and S.~Kaushik, S.,``Squeezing the local oscillator does not improve signal-to-noise ratio in heterodyne laser radar,'' Opt. Lett. {\bf 32}, 1369-1371 (2007).
\bibitem{Helstrom1976}C.~W.~Helstrom, {\em Quantum Detection and Estimation Theory}\/, (Academic Press,1976). 
\bibitem{Meyer1970}Meyer,~P.~L., {\em Introductory Probability and Statistical Applications}\/,
2nd ed., Addison-Wesley, Reading, MA, 1970.
\bibitem{GerryKnight2005}C.~G.~Gerry  and P.~L.~Knight,  {\em Introductory Quantum Optics}\/, (Cambridge University Press, 2005).
\bibitem{Haus2000} H.~A.~Haus,{\em Electromagnetic Noise and Quantum Optical Measurements}\/, (Springer, 2000).
\bibitem{GardinerZoller2004}C.~W.~Gardiner P.~Zoller, P. (2004), {\em Quantum Noise},\/ 3d ed., (Springer,2004).
\bibitem{Fabreetal2000}C.~Fabre, J.~B.~Fouet, and  A. ~Ma\^{i}tre, ``Quantum limits in the measurements of very small displacements
in optical images,'' Optics Letters {\bf 25}, 76-78 (2000).
\bibitem{Trepsetal2002}N.~Treps,  U.~Andersen, B.~Buchler, P.~K.~Lam,  A.~Ma\^{i}tre,  H.-A.~Bachor, and C.Fabre,
``Surpassing the standard quantum limit for optical imaging using nonclassical multimode light,''  Phys. Rev. Lett. {\bf 88},
203601 (2002).
\bibitem{SmithThomson1988} F.~G.~Smith and J.~H.~Thomson, {\em Optics}\/, 2nd ed. (John Wiley and Sons, 1988).
\bibitem{YuenChan1983}H.~P.~Yuen and V.~W.~S.~Chan, ``Noise in homodyne and heterodyne detection,'' Optics Letters {\bf 8}, 177-179 (1983).
\bibitem{AnnovazziLodietal1992}V.~Annovazzi-Lodi, S.~Donati,and S.~Merlo, ``Squeezed states in direct and coherent detection,'' Opt. Quan. Electron.
{\bf 24}, 285-310 (1992).



\end{thebibliography}
\end{document}